\documentclass[prd,aps,showpacs,,nofootinbib,superscriptaddress,preprintnumbers,epsf,psf]{revtex4}

\usepackage[english]{babel}
\usepackage{pstricks}
\usepackage[dvips]{graphicx}
\usepackage{epsf}
\usepackage{amsmath}
\usepackage{amsfonts}
\usepackage{amssymb}

\begin{document}

\begin{flushright}
CPHT-RR103.1111\\
LU TP 11-42\\
December 2011
\end{flushright}

\title{Gauge invariance of DVCS off
an arbitrary spin hadron: the deuteron target case}

\author{ I.~V.~Anikin}
\email{anikin@theor.jinr.ru, Igor.Anikin@physik.uni-regensburg.de}
\affiliation{Institute for Theoretical Physics, University of Regensburg,
             D-93040 Regensburg, Germany}
\affiliation{Bogoliubov Laboratory of Theoretical Physics, JINR,
             141980 Dubna, Russia}

\author{ R.~S.~Pasechnik}
 \email{Roman.Pasechnik@thep.lu.se}
 \affiliation{Department of Astronomy and Theoretical
 Physics, Lund University, SE 223-62 Lund, Sweden}

\author{ B.~Pire}
\email{pire@cpht.polytechnique.fr}
\affiliation{CPHT, \'Ecole Polytechnique, CNRS,
             91128 Palaiseau, France}

\author{ O.~V.~Teryaev}
\email{teryaev@theor.jinr.ru}
\affiliation{Bogoliubov Laboratory of Theoretical Physics, JINR,
             141980 Dubna, Russia}

\begin{abstract}
We study the deeply virtual Compton scattering off a spin-one
particle, as the case for the coherent scattering off a deuteron
target. We extend our approach, formulated initially for a spinless
case, and discuss the role of twist three contributions for
restoring the gauge invariance of the amplitude. Using twist three
contributions and relations, which emanate from the QCD equations of
motion, we derive the gauge invariant amplitude for the deeply
virtual Compton scattering (DVCS) off hadrons with spin $1$. Using
the derived gauge invariant amplitude, the single spin asymmetry is
discussed.
\end{abstract}
\pacs{12.38.Bx, 13.60.Le}
\date{\today}
\maketitle

\section{Introduction}

 Deeply virtual Compton scattering (DVCS) off the deu\-teron target
has recently attracted much attention from the experimental point of
view \cite{Ellinghaus,Mazouz,Voutier,HERMES-deuteron}. One of the
main reasons of this interest is the fact that the DVCS process
gives  information about the generalized parton distributions
(GPDs). From the theoretical point of view, the leading twist-2 GPDs
for the deuteron were studied in \cite{Cano,Berger,Kirch-Muel}.
However, the leading twist-2 accuracy for the DVCS amplitude,
calculated in the case where the final deuteron gets a significant
transverse momentum, is not enough for the study of such processes.
This is due to the fact that in the essential case of
sizable transverse
transfer momentum, $\Delta_T\neq 0$, the leading twist-2
approximation, in the Bjorken limit, is not sufficient for
the photon gauge invariance of the DVCS amplitude (see, for example, \cite{Gui98})
\footnote{Throughout this paper, we deal with the $U(1)$ gauge invariance of
the amplitude rather than the $SU(3)_c$ gauge invariance.
Concerning the QCD gauge, we fix it by choosing
the axial gauge for gluons, $A^+=0$.}.
Besides, the relevant terms are proportional to the transverse
component of the momentum transfer and provide the leading
contribution to some observables.

Extending the Ellis-Furmanski-Petronzio-Efremov-Teryaev (EFPET)
approach (see, \cite{E&F&P,ET}) to the non-forward case, this
problem was  first solved in \cite{APT} where it was demonstrated
how the inclusion of twist-3 contributions related to the matrix
elements of quark-antiquark-gluon operators, can restore the gauge
invariance of the DVCS amplitude  off a (pseudo)scalar particle
(e.g. pion, $He^4$). Then, the main ideas of \cite{APT} were used
and generalized for the nucleon target (see,
\cite{Pol-tw3,Bel-tw3,Bel2-tw3,Vander-tw3,Kivel-tw3,Rad-tw3} and
\cite{AIPSW10} for different processes).

In this paper, we adhere to the approach \cite{APT} and make a
comprehensive analysis of the twist three contributions to the DVCS
amplitude  off a spin-1 hadron\footnote{Actually, the
method described in \cite{APT} and in this paper is suitable for a
study of DVCS off an arbitrary spin hadron.}. Since the
parameterizations of the relevant hadronic matrix elements of the
quark-gluon operators depend much on the spin of the external
hadrons, following the very useful idea of \cite{Pol-tw3} we start
our study with a parametrization-free approach, and then apply it to
the specific case of spin-$1$ hadron.

\section{Kinematics and Approximations}
\label{kinematics}

\begin{figure*}[tb!]
\centerline{\includegraphics[width=0.35\textwidth]{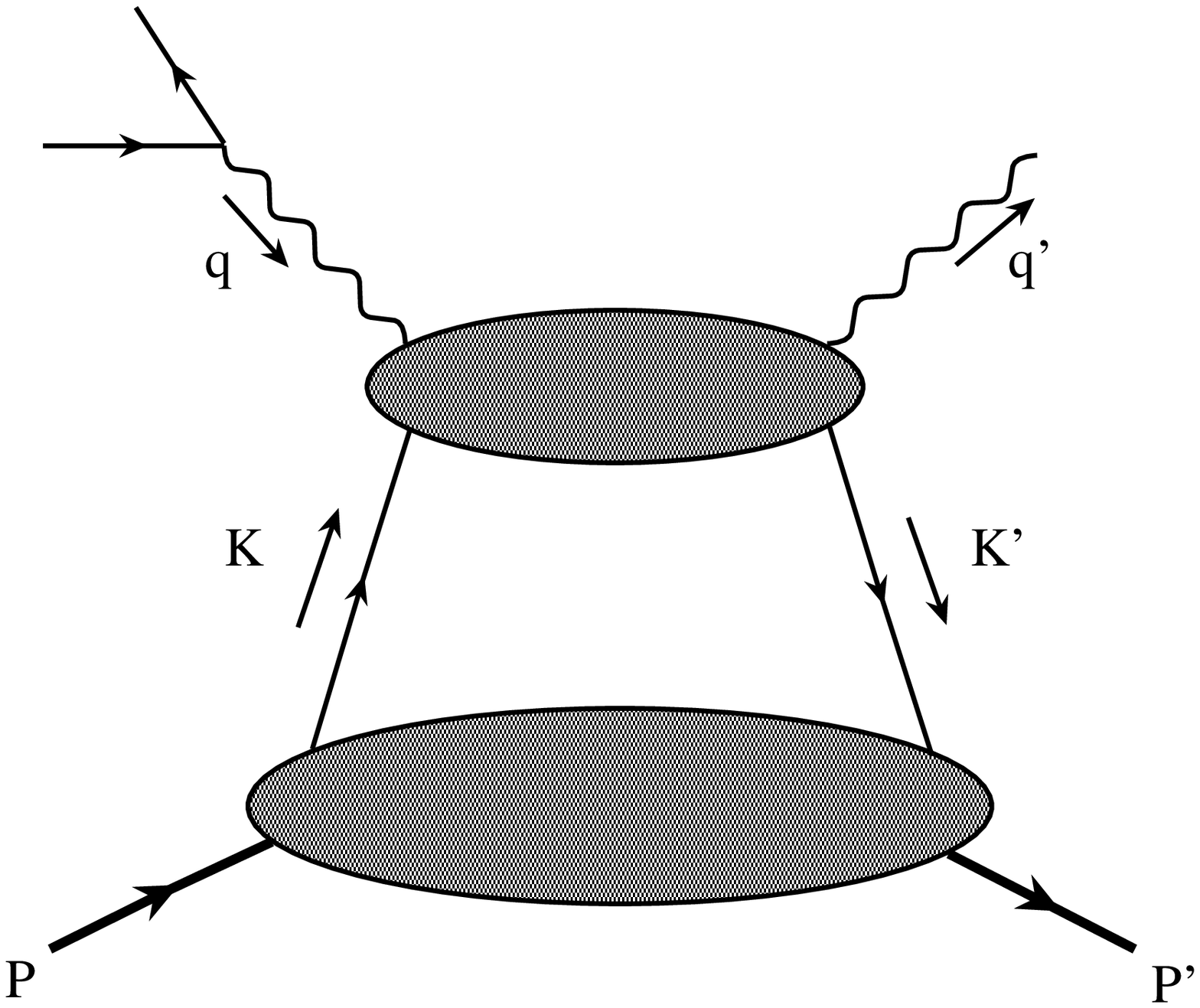}
\hspace{1.cm}\includegraphics[width=0.35\textwidth]{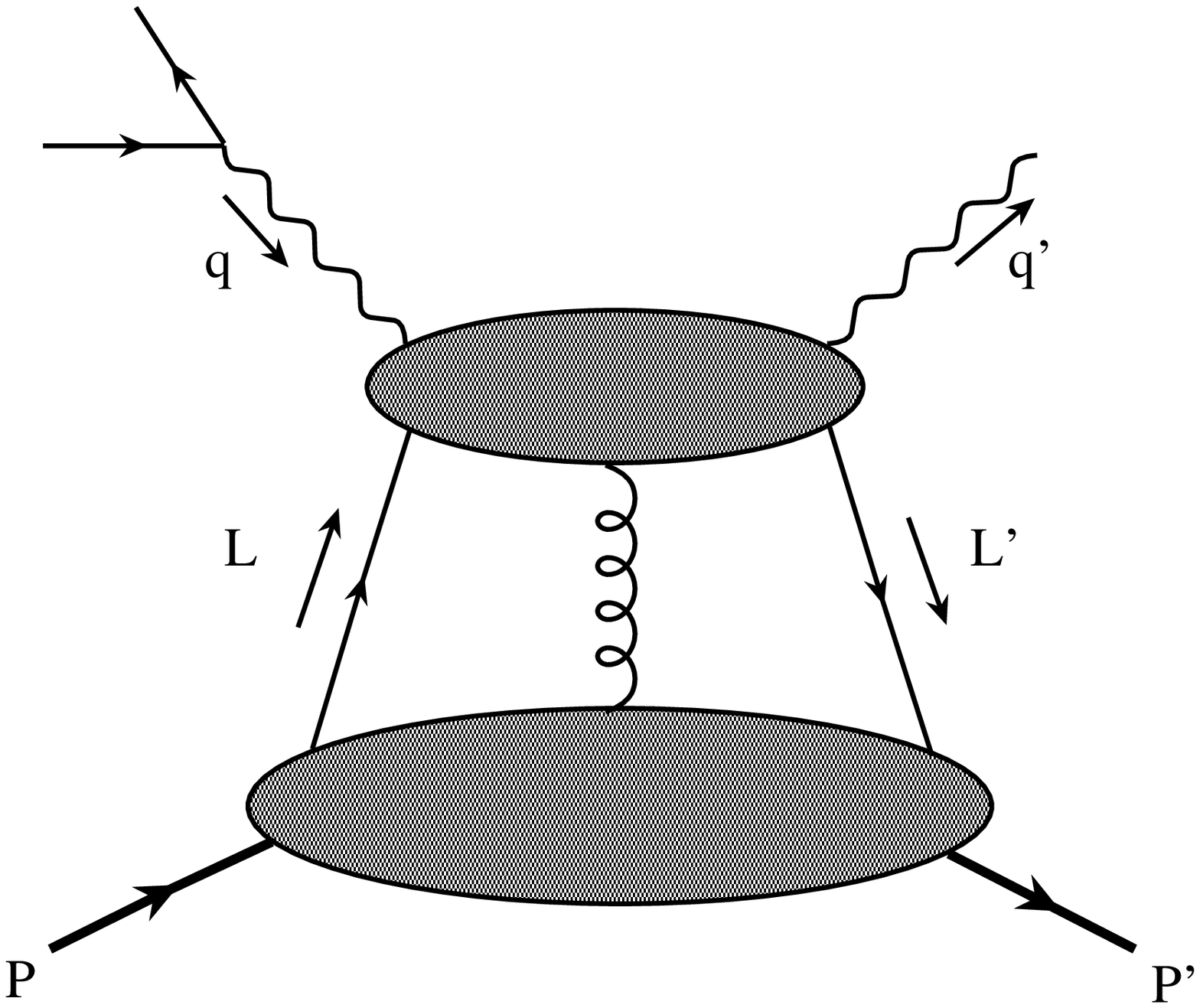}}
\vspace{1cm}
  \caption{The Feynman diagrams corresponding to the deeply virtual Compton scattering.
The DVCS amplitude with two-particle correlators is depicted on the
left panel, while the amplitude with the three-particle correlators
-- on the right panel. Notations: $P\equiv p_1,\quad
P^{\prime}\equiv p_2$, $K\equiv k-\Delta/2\approx xP-\Delta/2, \quad
K^{\prime}\equiv k+\Delta/2\approx xP+\Delta/2$, $L\equiv
k_1-\Delta/2\approx x_1P-\Delta/2, \quad L^\prime\equiv
k_2+\Delta/2\approx x_2P+\Delta/2$. Here, $k$ and $k_i$ correspond
to the loop momenta in the diagrams.} \label{Fig1}
\end{figure*}

Let us start with the discussion of the kinematics and approximations
which we use in this paper. The process we consider is
\begin{eqnarray}
\label{process}
\gamma^*(q) + D(p_1) \to \gamma(q^\prime) + D(p_2) .
\end{eqnarray}
Here, we mainly focus on the deuteron as a target but all our
approach is suitable for any spin-one hadron target. At the Born
level, the Feynman diagrams corresponding to the considered process
are depicted in Fig.~\ref{Fig1}. This process is a hard exclusive
reaction for which a QCD factorization theorem  applies. In
this case, the virtuality of the initial off-shell photon is used as the
 large scale, i.e. $q^2=-Q^2\to\infty$, while the final photon is
on-shell with $q^{\prime\, 2}=0$. Besides, this asymptotical
regime is identical to the light-cone formalism. Therefore, we first
introduce a  light-cone basis which is constructed by the
``plus'' and ``minus'' vectors:
\begin{eqnarray}\nonumber
&& n^\star=\Lambda(1,\, 0, \, 0,\, 1)\, , \\
&& n=\frac{1}{2\Lambda}(1,\, 0, \, 0,\, -1)\, , \label{lcb} \\
&& n^\star\cdot n =1\, ,\nonumber
\end{eqnarray}
where $\Lambda$ is an arbitrary and dimensionful constant which can
be expressed via the Lorentz invariants. The exact form of $\Lambda$
as a function of invariants depends on the frame which one works in.

In the present paper, we consider the DVCS amplitude up to the twist
three accuracy, discarding the contributions associated with the twist four and
higher. Such a constraint imposes the following
relations for the hadron average and transfer momenta :
\begin{eqnarray}
\label{kin}
&&P=\frac{p_1+p_2}{2}=n^\star+\frac{\bar M^2}{2}n\, \approx \, n^\star ,
\nonumber\\
&&\Delta=p_2-p_1 = -2\xi P + 2\xi\bar M^2 n  +\Delta^T \approx -2\xi P +\Delta^T,
\nonumber\\
&& P\cdot \Delta=0, \quad \Delta^2=t\approx 0 \, .
\end{eqnarray}
Notice that keeping the $\bar M^2$-term in the Sudakov decomposition
of the relative momentum $P$ (see, (\ref{kin})) leads to the
necessity to include the twist four contributions as well, which
goes beyond the scope of the present paper. Since corrections of the
order $O(\Delta_T^2/Q^2)$ demand a special care, at this moment, we
postpone the study of them until a forthcoming paper. Notice that
the detailed analysis of these contributions has recently been
presented in \cite{BM1,BM2}.

It is also instructive to introduce the photon average momentum:
\begin{eqnarray}
\nonumber
 &&\bar
Q=\frac{q+q^\prime}{2}=q-\frac{\Delta}{2}=q^\prime+\frac{\Delta}{2},
\quad q^\prime =(P\cdot q^\prime)\, n, \\
 &&(P\cdot
q^\prime)=(P\cdot \bar Q)= (P\cdot q) \, . \label{kin2}
\end{eqnarray}
One has to emphasize that the approximations discussed in this
section do not affect the generality of our study and can be applied
to a study of arbitrary spin hadrons.

\section{Factorization and the gauge invariant amplitude}

In this section, we briefly describe the factorization procedure
applied to the DVCS amplitude up to the twist three accuracy. The
details of this factorization can be found in
\cite{APT,AIPSW10,AT1,AT2}.

At the Born level, the sum of the amplitudes with the two- and
three-particle correlators, or $T_{\mu\nu}^{(1)}$ and
$T_{\mu\nu}^{(2)}$ amplitudes, has the following form (for the DIS
case, see \cite{ET}):
\begin{eqnarray}\nonumber
&&T_{\mu\nu}^{(1)}+T_{\mu\nu}^{(2)}= \int d^4 k \, \text{tr} \biggl[
E_{\mu\nu}(k) \Gamma (k) \biggr]\\&& + \int d^4 k_1 \, d^4 k_2 \,
\text{tr}\biggl[ E_{\mu\rho\nu}(k_1, k_2) \Gamma^{\rho} (k_1, k_2)
\biggr] \label{1}
\end{eqnarray}
where $E_{\mu\nu}$ and $E_{\mu\rho\nu}$ are the coefficient
functions, at the Born approximation, with two quark legs and two
quark and one gluon legs, respectively. In Eq.~(\ref{1}), we use the
following notations:
\begin{eqnarray}
&&\Gamma_{\underline{\alpha\beta}}(k)=-\int\limits_{-\infty}^{\infty}
d^4 z \, e^{i( k-\Delta/2) z } \, \langle p_2|
\psi_{\underline{\alpha}}(z) \bar\psi_{\underline{\beta}}(0)| p_1
\rangle ,
\nonumber\\
&&\Gamma^{\rho}_{\underline{\alpha\beta}} (k_1, k_2)=
-\int_{-\infty}^{\infty} d^4 z_1 \, d^4 z_2 \, e^{ i( k_1-\Delta/2)
z_1 +i(k_2-k_1) z_2 }\nonumber \\&& \times\langle p_2 |
\psi_{\underline{\alpha}}(z_1) g A^{\rho}(z_2)
\bar\psi_{\underline{\beta}}(0)| p_1 \rangle\, ,\label{2}
\end{eqnarray}
where the underlined indices $\underline{\alpha},\,
\underline{\beta}$ denote the Dirac spin indices while the other
indices correspond to the Lorentz ones.

It is convenient to choose the axial gauge condition for gluons,
{\it i.e.} $n\cdot A =0$, where $n$ is the light-cone vector defined
in Eq.~(\ref{lcb}). Then, we carry out a decomposition of the loop
momentum $k$ over the light-cone vectors (the Sudakov decomposition)
as follows
\begin{eqnarray}
\label{k} k_\mu =x\, P_\mu + (k\cdot P)n_\mu + k^T_\mu \approx x\,
P_\mu + k^T_\mu \, ,
\end{eqnarray}
where $x=k\cdot n$. As the next step of the factorization procedure,
we perform the replacement for the integration momentum in
Eq.~(\ref{1}) as
\begin{eqnarray}
\label{subst}
d^4k_i\, \to\, d^4k_i\, dx_i \delta(x_i-k_i\cdot n).
\end{eqnarray}
This allows us to expand the two-quark coefficient function
$E_{\mu\nu}$ (see, (\ref{1})) in a Taylor series:
\begin{eqnarray}
 \nonumber E_{\mu\nu}(k) &=& E_{\mu\nu}(x P) +
\frac{\partial E_{\mu\nu}(k)}{\partial k_\alpha}
\biggl|_{k=xP}\biggr. \, (k-x\, P)_\alpha + \ldots\,, \\&& (k-x \,
P)_\alpha \approx k^T_\alpha \,.\label{expand}
\end{eqnarray}
Then, using the collinear Ward identity (see, \cite{ET})
\begin{equation}
\frac{\partial E_{\mu\nu}(k)}{\partial k^\rho}=E_{\mu\rho\nu}(k,k) \, .
\end{equation}
we, finally, arrive at the factorized (in the momentum space) DVCS
amplitude which reads
\begin{eqnarray}
 \nonumber &&T_{\mu\nu}^{(1)}+T_{\mu\nu}^{(2)}=
\int\limits_{-1}^{1} dx \, \text{tr}\biggl[ E_{\mu\nu}(x P) \Gamma
(x) \biggr] \\&& +\, \int\limits_{-1}^{1} dx_1\, dx_2\,
\text{tr}\biggl[ E_{\mu\rho\nu}(x_1, x_2) \omega^{\rho\rho^{\prime}}
\Gamma_{\rho^{\prime}} (x_1, x_2) \biggr]\label{1.1}
\end{eqnarray}
where $\omega^{\rho\rho^{\prime}}=\delta^{\rho\rho^{\prime}}-
n^{\rho^{\prime}} P^{\rho}$,
and
\begin{eqnarray}
&&\Gamma_{\underline{\alpha\beta}}(x)=-\int\limits_{-\infty}^{\infty}
\frac{d\lambda}{2\pi} \, e^{ i( x+\xi)\lambda }\, \langle p_2 |
\psi_{\underline{\alpha}}(\lambda n)
\bar\psi_{\underline{\beta}}(0)| p_1 \rangle ,
\nonumber\\
&&\Gamma^{\rho^{\prime}}_{\underline{\alpha\beta}} (x_1, x_2)=
-\int\limits_{-\infty}^{\infty} \frac{d\lambda_1}{2\pi} \,
\frac{d\lambda_2}{2\pi} \, e^{ i( x_1+\xi) \lambda_1 +i(x_2-x_1)
\lambda_2 } \nonumber\\ &&\times\, \langle p_2 |
\psi_{\underline{\alpha}}(\lambda_1 n) \stackrel{ \leftrightarrow }{
iD^{\rho^{\prime}} }(\lambda_2 n) \bar\psi_{\underline{\beta}}(0) |
p_1 \rangle\, ,\label{2.1}
\end{eqnarray}
where
$\stackrel{\rightarrow}{iD^{\mu}}=\stackrel{\rightarrow}{i\partial^{\mu}}+gA^{\mu}$
is the QCD covariant derivative in the fundamental representation.
This amplitude is also needed to be ``factorized`` in the Dirac
space. This can be reached by making use of the Fierz decomposition
over spinor indices.

In fact, the contributions of $T^{(2)}_{\mu\nu}$ are not completely
independent from one another because of the QCD equations of motion (e.o.m.) for
fermions. The next step is to use the QCD e.o.m. in order to
reexpress the contributions of the correlators with the covariant
derivative through the correlators which include only $\bar\psi$ and
$\psi$ fields. Afterwards, the contribution of $T^{(2)}_{\mu\nu}$
presented in terms of the two-particle correlators should be
combined together with the contribution of $T^{(1)}_{\mu\nu}$ in
order to get the gauge invariant DVCS amplitude at the Born level.

Let us now focus on the QCD equations of motion. For the sake of
simplicity, we start within the approximation where the
three-particle correlators are absent.
Indeed, to derive the
gauge invariant amplitude, it is sufficient to consider only the
kinematical twist contributions since the kinematical and dynamical
twists enter in the QCD e.o.m. and the DVCS amplitude additively
(see \cite{APT,AT1,AT2,Kivel-tw3}).

Let us consider the e.o.m. in the following form:
\begin{eqnarray}
\label{eom}
\langle \biggl(\stackrel{\rightarrow}{i\hat\partial} \psi(z)\biggr) \bar\psi(0)\rangle=0 \, , \quad
\langle \psi(z) \biggl(\bar\psi(0)\stackrel{\leftarrow}{i\hat\partial}\biggr) \rangle=0 \, ,
\end{eqnarray}
where the Dirac spinor indices are omitted for simplicity.

We want to adapt our approach and use a parametri\-zation-free
formalism, according to \cite{Pol-tw3}. To this end, we introduce
the notations (here, $\Gamma$ denotes different combinations of
$\gamma$-matrices),
\begin{eqnarray}
\nonumber &&\langle p_2|\bar\psi(0) \Gamma \psi(z)|p_1\rangle
\stackrel{F}{=} {\cal F}^{[\Gamma]}(x)\,, \\ &&\langle
p_2|\bar\psi(0) \Gamma
\stackrel{\leftrightarrow}{i\partial^T_{\alpha}}\psi(z) |p_1\rangle
\stackrel{F}{=} {\cal
F}^{[\stackrel{\leftrightarrow}{\partial^T\,}\Gamma]}_{\alpha}(x) \,
,\label{par1}
\end{eqnarray}
where $\stackrel{F}{=}$ denotes the Fourier transformation with the
measure ($z=\lambda n, z^{\prime}=0 $)
\begin{eqnarray}
 dx e^{ -i( xP-\frac{\Delta}{2})z
+i( xP+\frac{\Delta}{2})z^{\prime} }.
\end{eqnarray}
In Eq.~(\ref{par1}), if $\Gamma$ becomes the $\gamma$-matrix with an
open Lorentz index, the functions ${\cal F}^{[\Gamma]}(x)$ and
${\cal
F}^{[\stackrel{\leftrightarrow}{\partial^T\,}\Gamma]}_{\alpha}(x)$
should be written with additional Lorentz indices.

Keeping the vector and axial-vector projections in the Fierz
decomposition of Eq.~(\ref{eom}) (all other structures do not
contribute in the massless quark case), the e.o.m., in terms of the
functions (\ref{par1}), take the following form
\begin{eqnarray}
\label{eom2}
 && \gamma^\alpha_T \gamma^- \biggl\{ {\cal
F}^{[\stackrel{\leftrightarrow}{\partial^T\,}\gamma^+]}_{\alpha}(x)
- x P^+ \,{\cal F}^{[\gamma_T]}_{\alpha}(x) \\
 && +\, \frac{i}{2}\varepsilon^{\Delta^T - \alpha +}\,{\cal
F}^{[\gamma^+\gamma_5]}(x) + \xi P^+ \, i \varepsilon^{\beta -
\alpha +}\,{\cal F}^{[\gamma_T\gamma_5]}_{\beta}(x)
\biggr\}=0 \, ,\nonumber \\
\label{eom3}
 && \gamma^\alpha_T \gamma^- \biggl\{ i
\varepsilon^{\beta - \alpha +}\, {\cal
F}^{[\stackrel{\leftrightarrow}{\partial^T\,}\gamma^+
\gamma_5]}_{\beta}(x) + \frac{\Delta_\alpha^T}{2} {\cal
F}^{[\gamma^+]}(x) \\
 && - x P^+ \, i\varepsilon^{\beta - \alpha
+}\,{\cal F}^{[\gamma_T\gamma_5]}_{\beta}(x) + \xi P^+ \,{\cal
F}^{[\gamma_T]}_{\alpha}(x) \biggr\}=0 \, . \nonumber
\end{eqnarray}
Following \cite{APT} and using Eq.~(\ref{par1}), the DVCS
amplitudes (see, Eq.~(\ref{1.1})) can be written as
\begin{eqnarray*}
 && T^{(1)}_{\mu\nu}=\int dx \,{\rm tr} \biggl[ \gamma_\nu \frac{x\hat
P + \hat{\bar Q}}{(x P + \bar Q)^2} \gamma_\mu \gamma^- \biggr]\,
{\cal F}^{[\gamma^+]}(x) + \\
 && \int dx \,{\rm tr} \biggl[
\gamma_\nu \frac{x\hat P + \hat{\bar Q}}{(x P + \bar Q)^2}
\gamma_\mu \gamma^T_\alpha \biggr]\, {\cal F}^{[\gamma_T]}_\alpha(x)
-
\nonumber\\
&& \int dx \,{\rm tr} \biggl[ \gamma_\nu \frac{x\hat P + \hat{\bar
Q}}{(x P + \bar Q)^2} \gamma_\mu \gamma^- \gamma_5 \biggr]\, {\cal
F}^{[\gamma^+\gamma_5]}(x) - \\
 && \int dx \,{\rm tr} \biggl[
\gamma_\nu \frac{x\hat P + \hat{\bar Q}}{(x P + \bar Q)^2}
\gamma_\mu \gamma^T_\alpha \gamma_5 \biggr]\, {\cal
F}^{[\gamma_T\gamma_5]}_\alpha(x)
\nonumber\\
&& +\, {\rm ``crossed"} \, ,
\end{eqnarray*}
and
\begin{eqnarray*}
&&T^{(2)}_{\mu\nu}= -\int dx \,{\cal
F}^{[\stackrel{\leftrightarrow}{\partial^T\,}\gamma^+]}_{\alpha}(x)\\
&& \times\, {\rm tr} \biggl[ \gamma_\nu \frac{x\hat P + \hat{\bar
Q}}{(x P + \bar Q)^2} \gamma^T_\alpha \frac{x\hat P + \hat{\bar
Q}}{(x P + \bar Q)^2} \gamma_\mu \gamma^- \biggr]  +
\nonumber\\
&&\int dx \, {\cal
F}^{[\stackrel{\leftrightarrow}{\partial^T\,}\gamma^+\gamma_5]}_{\alpha}(x)\\
&&\times\, {\rm tr} \biggl[ \gamma_\nu \frac{x\hat P + \hat{\bar
Q}}{(x P + \bar Q)^2} \gamma^T_\alpha \frac{x\hat P + \hat{\bar
Q}}{(x P + \bar Q)^2} \gamma_\mu \gamma^- \gamma_5 \biggr]
\nonumber\\
&& +\, {\rm ``crossed"} \, .
\end{eqnarray*}
As was mentioned above, we now have to use the QCD e.o.m., written
in the form of Eqs.~(\ref{eom2}) and (\ref{eom3}), for the amplitude
$T^{(2)}_{\mu\nu}$, and then, combining it with $T^{(1)}_{\mu\nu}$,
to collect all the similar terms in the final expression. Due to the
specific structure of the e.o.m., the correlators with transverse
derivatives in $T^{(2)}_{\mu\nu}$ can be eliminated, and can be
re-expressed through known correlators without derivatives (this
will also be valid for the case with the dynamical twist contributions included).
So, one gets
\begin{eqnarray}
\nonumber &&T^{(1)+(2)}_{\mu\nu}= \frac{1}{2P\cdot q} \int dx\,
\biggl( \frac{1}{x-\xi+i\epsilon} + \frac{1}{x+\xi-i\epsilon}
\biggr) {\cal T}_{\mu\nu}, \\&& \label{GI-ampl-F0}
\end{eqnarray}
where
\begin{eqnarray}
\nonumber
 && {\cal T}_{\mu\nu} = \biggl[ \xi ( \delta^\nu_+ P_{\mu}
+ \delta^\mu_+ P_{\nu})  + \delta^{\mu}_+ \bar Q_{\nu} +
\delta^{\nu}_+ \bar Q_{\mu} - g_{\mu\nu}\bar Q^-\\
 && +\, \frac{1}{2} \delta^{\mu}_+ \Delta_{\nu}^{T} - \frac{1}{2}
\delta^{\nu}_+ \Delta_{\mu}^{T} \biggr] \, {\cal F}^{[\gamma^+]}(x)
\nonumber\\
 && +\, \biggl[ 3\xi P_\mu g_{\nu\alpha}^T + \xi P_\nu g_{\mu\alpha}^T +
\bar Q_{\mu}g_{\nu\alpha}^T + \bar Q_{\nu}g_{\mu\alpha}^T  \biggr]\,
{\cal F}^{[\gamma_T]}_\alpha(x) \nonumber\\
 && +\, i\frac{\xi}{x}\biggl[ \Delta^T_\beta \,\delta^{\nu}_+  - \bar Q ^-
g^T_{\beta\nu} \biggr] \,\varepsilon^{\beta - \mu +}\,{\cal
F}^{[\gamma^+\gamma_5]}(x)
\nonumber\\
 && +\, i \frac{\xi}{x}\biggl[ -3\xi P^{\mu} g_{\beta\nu}^{T} + \xi
P^{\nu} g_{\beta\mu}^{T} +\bar Q^\nu g_{\mu\beta}^T - \bar Q^\mu
g_{\nu\beta}^T) \biggr]\nonumber \\
 && \times\, \varepsilon^{\alpha -
\beta +} {\cal F}^{[\gamma_T\gamma_5]}_{\alpha}(x)  \,
,\label{GI-ampl-F}
\end{eqnarray}
where $\delta^\nu_+$ denotes the usual Kronecker symbol.

We have thus derived the gauge invariant DVCS amplitude for the most
general case of a target with an arbitrary spin which totally
coincides with the results obtained in \cite{Pol-tw3} by a different
approach. In the present paper, the corrections of the order
$O(\Delta_T^2/Q^2)$ have been neglected. We postpone the discussion
of these corrections until a forthcoming analysis. The detailed
study of these contributions has recently been presented in
\cite{BM1,BM2}.

If one now specifies spin of the hadron and, then, uses explicit
parameterizations for the corresponding matrix elements, it will, in
particular, reproduce the known cases of spin-$0$ and spin-$1/2$
(see, \cite{APT}, \cite{Pol-tw3} -- \cite{Rad-tw3}).

We now study  DVCS off a spin-$1$ particle, which is of
phenomenological interest in the deuteron case \cite{CP,Berger}. To
this end, we first specify the parametrization of the relevant
matrix elements. Namely, the parameterizations for the vector
correlators at the leading twist-$2$ level are (see, \cite{Berger})
\begin{eqnarray}
\nonumber &&\langle p_2,\lambda_2| \left[ \bar\psi(0)\gamma_{\mu}
\psi(z) \right]^{\text{tw-2}} | p_1,\lambda_1 \rangle
\stackrel{F}{=}  {\cal F}^{[\gamma^+]}_{\mu}(x) =\\&&
e^{*}_{2\,\alpha} \, {\cal
V}^{(i),\,L}_{\alpha\beta,\,\mu}(n^\star,n,\Delta_T) e_{1\, \beta}\,
H^{V}_i(x,\xi)\,,\label{VparTw2}
\end{eqnarray}
where
\begin{eqnarray}
\nonumber
 && e^{*}_{2\,\alpha} \, {\cal V}^{(i),\,
L}_{\alpha\beta,\,\mu}(n^\star,n,\Delta_T) e_{1\, \beta} \,
H^{V}_i(x,\xi)=\\ \nonumber
 && P_{\mu}\, H^V_{1,..,5}(e^{*}_2, e_1; x,\xi)
\equiv P_{\mu}\, \biggl\{(e^{*}_2\cdot e_1)\, H^{V}_1(x,\xi)\\
 && +\, (e^{*}_2\cdot P)(e_1\cdot n)\, H^{V}_2(x,\xi) + (e^{*}_2\cdot
n)(e_1\cdot P)\, H^{V}_3(x,\xi) \biggr. \nonumber\\ \biggl.
 && +\, \frac{1}{M^2}(e^*_2\cdot P)(e_1\cdot P)\,
H^{V}_4(x,\xi) \nonumber \\
 && +\, M^2 (e^*_2\cdot n)(e_1\cdot n)\,
H^{V}_5(x,\xi) \biggr\}.\label{F}
\end{eqnarray}
Here, for the sake of conciseness, a new compact notation
$H^V_{1,..,5}(e^{*}_2, e_1; x,\xi)$ has been introduced. Now, we
are in a position to discuss the twist-$3$ operator matrix elements
and their parameterizations. For the vector quark correlator we have
\begin{eqnarray}
\nonumber &&\langle p_2,\lambda_2| \left[ \bar\psi(0)\gamma_{\mu}
\psi(z) \right]^{\text{tw-3}} | p_1 ,\lambda_1 \rangle
\stackrel{F}{=} {\cal F}^{[\gamma_T]}_\mu(x)=\\&& e^{*}_{2\,\alpha}
\, {\cal V}^{(i)\,T}_{\alpha\beta,\,\mu}(n^\star,n,\Delta_T) e_{1\,
\beta}\,G^{V}_i(x,\xi)\,,\label{VparTw3}
\end{eqnarray}
where
\begin{eqnarray}
\nonumber &&e^{*}_{2\,\alpha} \, {\cal
V}^{(i)\,T}_{\alpha\beta,\,\mu}(n^\star,n,\Delta_T) e_{1\,
\beta}\,G^{V}_i(x,\xi)=\\
 && \Delta^T_{\mu}\,
G^{V}_{1,..,5}(e^{*}_2, e_1; x,\xi) + e^{*\,T}_{2\,\mu} (e_1\cdot
P)\, G^{V}_6(x,\xi)
\nonumber\\
&& +\, e^{T}_{1\,\mu} (e^*_2\cdot P)\, G^{V}_7(x,\xi) +M^2\, e^{*\,
T}_{2\, \mu} (e_1\cdot n)\, G^{V}_8(x,\xi) \nonumber\\
&& +\, M^2\, e^{T}_{1\, \mu} (e^*_2\cdot n)\, G^{V}_9(x,\xi)\,
.\label{tw3V}
\end{eqnarray}
Our next step is the parametrization of the axial-vector
correlator. In contrast to the vector projection, the Schouten
identity plays a crucial role in the determination of the Lorentz
independent structures. The twist-$2$ axial-vector correlator can be
parameterized by
\begin{eqnarray}
\nonumber &&\langle p_2,\lambda_2| \left[
\bar\psi(0)\gamma_{\mu}\gamma_5 \psi(z) \right]^{\text{tw-2}} | p_1
,\lambda_1 \rangle \stackrel{F}{=} {\cal
F}^{[\gamma^+\gamma_5]}_{\mu}(x) =\\
 && -i\, e^{*}_{2\,\alpha} \,
{\cal A}^{(i),\,L}_{\alpha\beta,\,\mu}(n^\star,n,\Delta_T) e_{1\,
\beta}\, H^{A}_i(x,\xi)\, ,\label{AparTw2}
\end{eqnarray}
where
\begin{eqnarray}\nonumber
&&e^{*}_{2\,\alpha} \, {\cal
A}^{(i),\,L}_{\alpha\beta,\,\mu}(n^\star,n,\Delta_T) e_{1\, \beta}\,
H^{A}_i(x,\xi)=\\&& \varepsilon_{\mu P e^{*\,T}_2 e^{T}_1}\,
H^A_1(x,\xi) + \frac{1}{M^2}\, \varepsilon_{\mu P \Delta^T
e^{*\,T}_2} (e_1\cdot P)\, H^A_2(x,\xi)
\nonumber\\
 && + \, \frac{1}{M^2}\,\varepsilon_{\mu P \Delta^T e^T_1}
(e^*_2\cdot P)\, H^A_3(x,\xi) \nonumber \\&&
 +\, \varepsilon_{\mu P \Delta^T
e^{*\,T}_2} (e_1\cdot n)\, H^A_4(x,\xi)\, . \nonumber
\end{eqnarray}

Next, let us consider the twist-$3$ correlators. Using the
light-cone basis, we have fifteen different possible tensors. As in
the twist-$2$ case, the use of the Schouten identity reduces the
number of independent tensors. Indeed, instead of fifteen possible
structures we have nine independent tensors which parameterize the
twist-3 axial-vector correlators. Finally, the axial-vector
correlator reads
\begin{eqnarray}
\nonumber &&\langle p_2,\lambda_2| \left[
\bar\psi(0)\gamma_{\mu}\gamma_5 \psi(z) \right]^{\text{tw-3}} | p_1
,\lambda_1 \rangle \stackrel{F}{=}{\cal
F}^{[\gamma_T\gamma_5]}_{\mu}(x) =\\&& -i e^{*}_{2\,\alpha} \, {\cal
A}^{(i)\,T}_{\alpha\beta,\,\mu}(n^\star,n,\Delta_T) e_{1\,
\beta}\,G^{A}_i(x,\xi)\,,\label{AparTw3}
\end{eqnarray}
where
\begin{eqnarray*}
&&e^{*}_{2\,\alpha} \, {\cal
A}^{(i)\,T}_{\alpha\beta,\,\mu}(n^\star,n,\Delta_T) e_{1\,
\beta}\,G^{A}_i(x,\xi)=\\
 && \varepsilon_{\mu n P e_1^T}(e^*_2\cdot
P)\, G^A_1(x,\xi) + \varepsilon_{\mu n P e^{*\,T}_2} (e_1\cdot P)\,
G^A_2(x,\xi)
\nonumber\\
&& + \, M^2\, \varepsilon_{\mu n P e^T_1} (e^*_2\cdot n)\,
G^A_3(x,\xi) + M^2\, \varepsilon_{\mu n P e^{*\, T}_2} (e_1\cdot
n)\,\\&& \times \, G^A_4(x,\xi) + \frac{1}{M^2}\,\varepsilon_{\mu
\Delta_T P e^*_2} (e_1\cdot P)\, G^A_5(x,\xi)
\nonumber\\
&& +\, \varepsilon_{\mu \Delta_T P e^*_2} (e_1\cdot n)\,
G^A_6(x,\xi) +
 \varepsilon_{\mu \Delta_T P e_1} (e^*_2\cdot n) \\
&& \times\, G^A_7(x,\xi) + \varepsilon_{\mu \Delta_T n e^{*}_2}
(e_1\cdot P)\, G^A_8(x,\xi)
\nonumber\\
&& +\, M^2\, \varepsilon_{\mu \Delta_T n e_1} (e^*_2\cdot n)\,
G^A_9(x,\xi) .
\end{eqnarray*}

Inserting the explicit parameterizations (\ref{VparTw2}) --
(\ref{AparTw3}) into the amplitude (\ref{GI-ampl-F}), we derive the
gauge invariant DVCS amplitude for the case of deuteron target:
\begin{eqnarray}
\nonumber && T_{\mu\nu}^{(\lambda_1,\, \lambda_2)} =
\frac{1}{2P\cdot \bar Q}\int dx \frac{1}{x-\xi+i\epsilon}
\\ && \times\, \Biggl({\cal T}^{(1)}_{\mu\nu}+{\cal T}^{(2)}_{\mu\nu}
+{\cal T}^{(3)}_{\mu\nu} +{\cal
T}^{(4)}_{\mu\nu}\Biggr)^{(\lambda_1,\, \lambda_2)} \nonumber
\\ && +\, O(\Delta^2_T;\, \bar M^2) + {\rm ``crossed"}\, ,\label{amp}
\end{eqnarray}
where the structure amplitudes ${\cal T}^{(k)}_{\mu\nu}$ read
\begin{eqnarray}
\nonumber &&{\cal T}^{(1)}_{\mu\nu} = H^V_{1,..,4}(x; e_1, e^*_2)
\Biggl( 2\xi P_{\mu}P_{\nu}  + P_{\mu}\bar Q_{\nu} + P_{\nu}\bar
Q_{\mu}
\\&& -\, g_{\mu\nu}(P\cdot \bar Q) +
\frac{1}{2}P_{\mu}\Delta_{\nu}^{T} -
\frac{1}{2}P_{\nu}\Delta_{\mu}^{T} \Biggr) + G^V_{1,..,4}(x; e_1,
e^*_2) \nonumber \\&& \times\, \Biggl( \xi P_{\nu}\Delta_{\mu}^T +
3\xi P_{\mu}\Delta_{\nu}^T + \Delta_{\mu}^{T}\bar Q_{\nu} +
\Delta_{\nu}^{T} \bar Q_{\mu} \Biggr)
\nonumber\\
&&
-\, \Biggl(\frac{(e^*_2\cdot P)(e_1\cdot P)}{M^2}G^A_5(x)+
(e^*_2\cdot P)(e_1\cdot n)G^A_6(x) \Biggr.
\nonumber\\
\Biggl. && +\, (e_1\cdot P)(e^*_2\cdot n)\left(
G^A_7(x)-G^A_8(x)\right) \Biggr) \nonumber \\&&\times\, \Biggl( 3\xi
P_{\mu}\Delta_{\nu}^{T} - \xi P_{\nu}\Delta_{\mu}^{T} -
\Delta_{\mu}^{T}\bar Q_{\nu} + \Delta_{\nu}^{T}\bar Q_{\mu}
\Biggr)\, ,\label{T1}
\end{eqnarray}
and
\begin{eqnarray}
\nonumber && {\cal T}^{(2)}_{\mu\nu} = (e_1\cdot P)G^V_6(x) \Biggl(
\xi P_{\nu} e^{*\,T}_{2\, \mu} + 3\xi P_{\mu}e^{*\,T}_{2\, \nu} +
e^{*\,T}_{2\, \mu}\bar Q_{\nu} \Biggr. \\ \Biggl. && +\,
e^{*\,T}_{2\, \nu} \bar Q_{\mu} \Biggr)+(e_1\cdot P)G^A_2(x) \Biggl(
3\xi P_{\mu}e^{*\,T}_{2\, \nu} - \xi P_{\nu}e^{*\,T}_{2\, \mu}
\Biggr. \nonumber \\ \Biggl. && -\, e^{*\,T}_{2\, \mu} \bar Q_{\nu}
+ e^{*\,T}_{2\, \nu} \bar Q_{\mu} \Biggr)\, , \label{T2}
\end{eqnarray}
and
\begin{eqnarray}\nonumber
&& {\cal T}^{(3)}_{\mu\nu} = (e^{*}_2\cdot P)G^V_7(x) \Biggl( \xi
P_{\nu} e^{T}_{1\, \mu} + 3\xi P_{\mu}e^{T}_{1\, \nu} + e^{T}_{1\,
\mu}\bar Q_{\nu} \Biggr. \\ \Biggl. && +\, e^{T}_{1\, \nu} \bar
Q_{\mu} \Biggr) + (e^{*}_2\cdot P)G^A_1(x) \Biggl( 3\xi
P_{\mu}e^{T}_{1\, \nu} - \xi P_{\nu}e^{T}_{1\, \mu} \Biggr. \nonumber \\
\Biggl. && -\, e^{T}_{1\, \mu} \bar Q_{\nu} + e^{T}_{1\, \nu} \bar
Q_{\mu} \Biggr)\, ,\label{T3}
\end{eqnarray}
and
\begin{eqnarray}
\nonumber && {\cal T}^{(4)}_{\mu\nu} = \varepsilon_{\mu\nu P n}
\Biggl( \varepsilon_{n P e^{*\,T}_2 e^{T}_1}\, H^A_1(x,\xi) \Biggr.
\nonumber\\
\Biggl. && +\, \frac{1}{M^2}\, \varepsilon_{n P \Delta^T e^{*\,T}_2}
(e_1\cdot P)\, H^A_2(x,\xi) \Biggr.
\nonumber\\
\Biggl. && +\, \frac{1}{M^2}\,\varepsilon_{n P \Delta^T e^T_1}
(e^*_2\cdot P)\, H^A_3(x,\xi) \Biggr.
\nonumber\\
\Biggl. && +\, \varepsilon_{n P \Delta^T e^{*\,T}_2} (e_1\cdot n)\,
H^A_4(x,\xi) \Biggr) \,. \label{T4}
\end{eqnarray}
This gauge invariant amplitude for  DVCS off deuteron is our main
result. For the sake of brevity, in Eqs.~(\ref{T1}) -- (\ref{T3}),
we neglected all terms which are proportional to the square of the
hadron mass. The full expressions for all amplitudes will be
presented in our forthcoming study.

\section{Single Spin Asymmetry}

In the preceding section, we have obtained the gauge invariant DVCS
amplitude which has a significant meaning for the investigation of
any observables. As a phenomenologically important example, we now
consider the single (electron) spin asymmetry (SSA), which arises in
the collision of the longitudinally polarized electron beams with an
unpolarized hadron target. The SSA parameter is defined as
\begin{eqnarray}
\label{ssa}
{\cal A}_L=\frac{d\sigma(\rightarrow)-d\sigma(\leftarrow)}
{d\sigma(\rightarrow)+d\sigma(\leftarrow)}.
\end{eqnarray}
The numerator of Eq.~(\ref{ssa}) can be expressed through the
imaginary part, first, of the interference between the twist-$2$ and
twist-$3$ helicity DVCS amplitudes and, second, of the interference
between the Bethe-Heitler (BH) and DVCS amplitudes.
For the JLAB kinematics, the $|{\cal A}_{\mathrm{DVCS}}|^2$
contribution can be neglected compared to the interference term because of
large contribution of the BH amplitude.

The DVCS amplitude contributing to exclusive real photon production
at $Q^2\gg M^2$ for the real and virtual photon polarizations, $i$
and $j$, reads
\begin{eqnarray}\nonumber
&& {\cal A}^{(i)}_{\rm{DVCS}}=\frac{e_\ell
e_q^2}{q^2}\sum_jL^{(j)}{\cal A}_{(j,i)}\,,\\
&& L^{(j)}={\cal
L}_{\mu'}(\ell_1,\ell_2){\epsilon^*_{\mu'}}^{(j)}\,,
\label{DVCS}
\end{eqnarray}
respectively. Here, the helicity amplitude is given by
\begin{eqnarray}
{\cal
A}_{(j,i)}=\epsilon_{\mu}^{(j)}T_{\mu\nu}{\epsilon_{\nu}'}^{*(i)},\quad
i=\pm1,\quad j=0,\,\pm1\,.
\end{eqnarray}
The Bethe-Heitler amplitude reads
\begin{eqnarray}
\nonumber && {\cal A}_{BH}^{(i)}=\frac{e_\ell
e_q^2}{\Delta^2}\sum_j\Lambda^{(j,i)}{\cal T}_{(j)},\quad {\cal
T}_{(j)}=\epsilon_{\mu}^{(j)}F_{\mu}\,,\\ &&
\Lambda^{(j,i)}=L_{\mu'\nu'}(\ell_1,\ell_2){\epsilon^*_{\mu'}}^{(j)}
{\epsilon_{\nu'}'}^{*(i)} \, ,\label{BH}
\end{eqnarray}
where
\begin{eqnarray}
\Delta^2=-4\xi^2\bar M^2+\Delta_T^2\equiv t \, ,
\end{eqnarray}
where $t$ is negative,
For convenience, we introduce the following shorthand notations for
Compton form factors related to the various GPDs:
\begin{eqnarray*}
&&\int dx \frac{G_i^V(x,\xi)}{x-\xi+i\epsilon} \Longrightarrow {\cal
G}_i^V,\quad \int dx
\frac{H_i^V(x,\xi)}{x-\xi+i\epsilon}\Longrightarrow {\cal {\cal
H}}_i^V,\\ && \int
dx\,\frac{\xi}{x}\,\frac{G_i^A(x,\xi)}{x-\xi+i\epsilon}\Longrightarrow
{\cal G}_i^A \, ,
\end{eqnarray*}
and
\begin{eqnarray*}
&& \int dx \frac{G_i^V(x,\xi)}{x-\xi-i\epsilon} \Longrightarrow
\overline{{\cal G}_i^V},\quad \int dx
\frac{H_i^V(x,\xi)}{x-\xi-i\epsilon}\Longrightarrow \overline{{\cal
{\cal H}}_i^V},\\ && \int
dx\,\frac{\xi}{x}\,\frac{G_i^A(x,\xi)}{x-\xi-i\epsilon}\Longrightarrow
\overline{{\cal G}_i^A} \, ,
\end{eqnarray*}

We now calculate the contribution to ${\cal A}_{\mathrm{BH}}^*{\cal A}_{\mathrm{DVCS}}$ coming from
the interference between Eqs.~(\ref{DVCS}) and (\ref{BH}).
We have the
following expressions (here, deuteron polarizations are summed up):
\begin{eqnarray}
\frac{1}{q^2\Delta^2}\sum_i[L^{(0)}{\cal
A}_{(0,i)}]\cdot [\Lambda^{(+,i)}{\cal T}_{(+)}]^* \biggl.\biggr|_{\text{tw}-2}  \sim \frac{1}{\xi
(\rho -4) \rho}\sum_{i=1}^{5}\sum_{j=1}^{3} {\cal H}_i^V \, C_{ij}^{(1)}\,  G_j,\,
\end{eqnarray}
where
\begin{eqnarray}
 &&C_{ij}^{(1)}=
\nonumber\\
&&\tiny{
\left(
\begin{array}{ccc}
 -8 \left(\xi ^2 (-4+\rho )-\rho \right) (-4+\rho ) (12+(-4+\rho ) \rho ) & 8 \left(\xi ^2 (-4+\rho )-\rho \right) (-4+\rho ) (-2+\rho ) \rho  & -8 \left(\xi ^2 (-4+\rho )-\rho \right) (-4+\rho ) (-2+\rho ) \rho  \\
 -4 \left(\xi ^2 (-4+\rho )-\rho \right) (-4+\rho ) (-2+\rho ) (\xi  (-4+\rho )+\rho ) & 4 \left(\xi ^2 (-4+\rho )-\rho \right) (-4+\rho ) \rho  (-2+\xi  (-4+\rho )+\rho ) & -4 \left(\xi ^2 (-4+\rho )-\rho \right) (-4+\rho ) \rho  (\xi  (-4+\rho )+\rho ) \\
 4 (\xi  (-4+\rho )-\rho ) \left(\xi ^2 (-4+\rho )-\rho \right) (-4+\rho ) (-2+\rho ) & -4 (2+\xi  (-4+\rho )-\rho ) \left(\xi ^2 (-4+\rho )-\rho \right) (-4+\rho ) \rho  & 4 (\xi  (-4+\rho )-\rho ) \left(\xi ^2 (-4+\rho )-\rho \right) (-4+\rho ) \rho  \\
 \left(\xi ^2 (-4+\rho )-\rho \right) (-4+\rho )^2 (-2+\rho ) \rho  & -\left(\xi ^2 (-4+\rho )-\rho \right) (-4+\rho )^2 \rho ^2 & \left(\xi ^2 (-4+\rho )-\rho \right) (-4+\rho )^2 \rho ^2 \\
 -16 \left(-2+\xi ^2 (-6+\rho )-\rho \right) \left(\xi ^2 (-4+\rho )-\rho \right) (-4+\rho ) & 16 (-4+\rho ) \left(-\xi ^2 (-4+\rho )+\rho \right)^2 & -16 \left(\xi ^4 (-4+\rho )^3-2 \xi ^2 (-4+\rho ) (-2+\rho ) \rho +\rho ^3\right)
\end{array}
\right)
}\, ,
\nonumber
\end{eqnarray}

\begin{eqnarray}
\frac{1}{q^2\Delta^2}\sum_i[L^{(0)}{\cal
A}_{(0,i)}]\cdot [\Lambda^{(+,i)}{\cal T}_{(+)}]^* \biggl.\biggr|_{\text{tw}-3}^V  \sim \frac{1}{\xi
(\rho -4) \rho}\Biggl(\sum_{i=1}^{9}\sum_{j=1}^{2} {\cal G}_i^V \, C_{ij}^{(2,1)}\,  G_j +
\sum_{i=1}^{9} {\cal G}_i^V \, C_{i}^{(2,2)}\,  G_3\Biggr),\,
\end{eqnarray}
where
\begin{eqnarray}
 &&C_{ij}^{(2,1)}=
\nonumber\\
&&\tiny{
\left(
\begin{array}{cc}
 -16 \xi  \left(\xi ^2 (-4+\rho )-\rho \right) (-4+\rho ) (12+(-4+\rho ) \rho ) & 16 \xi  \left(\xi ^2 (-4+\rho )-\rho \right) (-4+\rho ) (-2+\rho ) \rho  \\
 -8 \xi  \left(\xi ^2 (-4+\rho )-\rho \right) (-4+\rho ) (-2+\rho ) (\xi  (-4+\rho )+\rho ) & 8 \xi  \left(\xi ^2 (-4+\rho )-\rho \right) (-4+\rho ) \rho  (-2+\xi  (-4+\rho )+\rho ) \\
 8 \xi  (\xi  (-4+\rho )-\rho ) \left(\xi ^2 (-4+\rho )-\rho \right) (-4+\rho ) (-2+\rho ) & -8 \xi  (2+\xi  (-4+\rho )-\rho ) \left(\xi ^2 (-4+\rho )-\rho \right) (-4+\rho ) \rho  \\
 2 \xi  \left(\xi ^2 (-4+\rho )-\rho \right) (-4+\rho )^2 (-2+\rho ) \rho  & -2 \xi  \left(\xi ^2 (-4+\rho )-\rho \right) (-4+\rho )^2 \rho ^2 \\
 -32 \xi  \left(-2+\xi ^2 (-6+\rho )-\rho \right) \left(\xi ^2 (-4+\rho )-\rho \right) (-4+\rho ) & 32 \xi  (-4+\rho ) \left(-\xi ^2 (-4+\rho )+\rho \right)^2 \\
 -4 \xi  \left(\xi ^2 (-4+\rho )-\rho \right) (-4+\rho )^2 (-2+\rho ) & 4 (-1+\xi ) \xi  (-4+\rho )^2 \rho  (\xi  (-4+\rho )+\rho ) \\
 4 \xi  \left(\xi ^2 (-4+\rho )-\rho \right) (-4+\rho )^2 (-2+\rho ) & -4 \xi  (1+\xi ) (\xi  (-4+\rho )-\rho ) (-4+\rho )^2 \rho  \\
 16 \xi  \left(\xi ^2 (-4+\rho )-\rho \right) (-4+\rho ) (-2+\xi  (-6+\rho )+\rho ) & -16 \xi  (-4+\rho ) \left(\xi ^3 (-4+\rho )^2-\xi  (-8+\rho ) \rho -(-2+\rho ) \rho +\xi ^2 (-4+\rho ) (2+\rho )\right) \\
 16 \xi  (2+\xi  (-6+\rho )-\rho ) \left(\xi ^2 (-4+\rho )-\rho \right) (-4+\rho ) & -16 \xi  (-4+\rho ) \left(\xi ^3 (-4+\rho )^2-\xi  (-8+\rho ) \rho +(-2+\rho ) \rho -\xi ^2 (-4+\rho ) (2+\rho )\right)
\end{array}
\right)
}\,,
\nonumber
\end{eqnarray}

\begin{eqnarray}
 C_{i}^{(2,2)}=
\tiny{
\left(
\begin{array}{c}
 -16 \xi  \left(\xi ^2 (-4+\rho )-\rho \right) (-4+\rho ) (-2+\rho ) \rho  \\
 -8 \xi  \left(\xi ^2 (-4+\rho )-\rho \right) (-4+\rho ) \rho  (\xi  (-4+\rho )+\rho ) \\
 8 \xi  (\xi  (-4+\rho )-\rho ) \left(\xi ^2 (-4+\rho )-\rho \right) (-4+\rho ) \rho  \\
 2 \xi  \left(\xi ^2 (-4+\rho )-\rho \right) (-4+\rho )^2 \rho ^2 \\
 -32 \xi  \left(\xi ^4 (-4+\rho )^3-2 \xi ^2 (-4+\rho ) (-2+\rho ) \rho +\rho ^3\right) \\
 -4 \xi  \left(\xi ^2 (-4+\rho )-\rho \right) (-4+\rho )^2 \rho  \\
 4 \xi  \left(\xi ^2 (-4+\rho )-\rho \right) (-4+\rho )^2 \rho  \\
 16 \xi  \left(\xi ^2 (-4+\rho )-\rho \right) (-4+\rho ) (\xi  (-4+\rho )+\rho ) \\
 16 \xi  (\xi  (-4+\rho )-\rho ) \left(\xi ^2 (-4+\rho )-\rho \right) (-4+\rho )
\end{array}
\right)
}\,,
\nonumber
\end{eqnarray}

\begin{eqnarray}
\frac{1}{q^2\Delta^2}\sum_i[L^{(0)}{\cal
A}_{(0,i)}]\cdot [\Lambda^{(+,i)}{\cal T}_{(+)}]^* \biggl.\biggr|_{\text{tw}-3}^A  \sim \frac{1}{\xi
(\rho -4) \rho}\Biggl(\sum_{i=1}^{9}\sum_{j=1}^{2} \overline{{\cal G}}_i^A \, C_{ij}^{(3,1)}\,  G_j +
\sum_{i=1}^{9} \overline{{\cal G}}_i^A \, C_{i}^{(3,2)}\,  G_3\Biggr),\,
\end{eqnarray}
where
\begin{eqnarray}
 &&C_{ij}^{(3,1)}=
\nonumber\\
&&\tiny{
\left(
\begin{array}{cc}
 4 \xi  \left(\xi ^2 (-4+\rho )-\rho \right) (-4+\rho )^2 (-2+\rho ) & -4 \xi  (1+\xi ) (\xi  (-4+\rho )-\rho ) (-4+\rho )^2 \rho  \\
 -4 \xi  \left(\xi ^2 (-4+\rho )-\rho \right) (-4+\rho )^2 (-2+\rho ) & 4 (-1+\xi ) \xi  (-4+\rho )^2 \rho  (\xi  (-4+\rho )+\rho ) \\
 16 \xi  (2+\xi  (-6+\rho )-\rho ) \left(\xi ^2 (-4+\rho )-\rho \right) (-4+\rho ) & -16 \xi  (-4+\rho ) \left(\xi ^3 (-4+\rho )^2-\xi  (-8+\rho ) \rho +(-2+\rho ) \rho -\xi ^2 (-4+\rho ) (2+\rho )\right) \\
 16 \xi  \left(\xi ^2 (-4+\rho )-\rho \right) (-4+\rho ) (-2+\xi  (-6+\rho )+\rho ) & -16 \xi  (-4+\rho ) \left(\xi ^3 (-4+\rho )^2-\xi  (-8+\rho ) \rho -(-2+\rho ) \rho +\xi ^2 (-4+\rho ) (2+\rho )\right) \\
 -2 \xi  \left(\xi ^2 (-4+\rho )-\rho \right) (-4+\rho )^2 (-2+\rho ) \rho  & 2 \xi  \left(\xi ^2 (-4+\rho )-\rho \right) (-4+\rho )^2 \rho ^2 \\
 8 \xi  \left(\xi ^2 (-4+\rho )-\rho \right) (-4+\rho ) (-2+\rho ) (\xi  (-4+\rho )+\rho ) & -8 \xi  \left(\xi ^2 (-4+\rho )-\rho \right) (-4+\rho ) \rho  (-2+\xi  (-4+\rho )+\rho ) \\
 -8 \xi  (\xi  (-4+\rho )-\rho ) \left(\xi ^2 (-4+\rho )-\rho \right) (-4+\rho ) (-2+\rho ) & 8 \xi  (2+\xi  (-4+\rho )-\rho ) \left(\xi ^2 (-4+\rho )-\rho \right) (-4+\rho ) \rho  \\
 8 \xi  (\xi  (-4+\rho )-\rho ) \left(\xi ^2 (-4+\rho )-\rho \right) (-4+\rho ) (-2+\rho ) & -8 \xi  (2+\xi  (-4+\rho )-\rho ) \left(\xi ^2 (-4+\rho )-\rho \right) (-4+\rho ) \rho  \\
 -32 \xi  \left(-2+\xi ^2 (-6+\rho )-\rho \right) \left(\xi ^2 (-4+\rho )-\rho \right) (-4+\rho ) & 32 \xi  (-4+\rho ) \left(-\xi ^2 (-4+\rho )+\rho \right)^2
\end{array}
\right)
}\,,
\nonumber
\end{eqnarray}

\begin{eqnarray}
C_{i}^{(3,2)}=
\tiny{
\left(
\begin{array}{c}
 4 \xi  \left(\xi ^2 (-4+\rho )-\rho \right) (-4+\rho )^2 \rho  \\
 -4 \xi  \left(\xi ^2 (-4+\rho )-\rho \right) (-4+\rho )^2 \rho  \\
 16 \xi  (\xi  (-4+\rho )-\rho ) \left(\xi ^2 (-4+\rho )-\rho \right) (-4+\rho ) \\
 16 \xi  \left(\xi ^2 (-4+\rho )-\rho \right) (-4+\rho ) (\xi  (-4+\rho )+\rho ) \\
 -2 \xi  \left(\xi ^2 (-4+\rho )-\rho \right) (-4+\rho )^2 \rho ^2 \\
 8 \xi  \left(\xi ^2 (-4+\rho )-\rho \right) (-4+\rho ) \rho  (\xi  (-4+\rho )+\rho ) \\
 -8 \xi  (\xi  (-4+\rho )-\rho ) \left(\xi ^2 (-4+\rho )-\rho \right) (-4+\rho ) \rho  \\
 8 \xi  (\xi  (-4+\rho )-\rho ) \left(\xi ^2 (-4+\rho )-\rho \right) (-4+\rho ) \rho  \\
 -32 \xi  \left(\xi ^4 (-4+\rho )^3-2 \xi ^2 (-4+\rho ) (-2+\rho ) \rho +\rho ^3\right)
\end{array}
\right)
}\,.
\nonumber
\end{eqnarray}
Here, $\rho = t/M^2$. In the last expression, $G_{1,2,3}$ are
the known electromagnetic deuteron form factors.

As it can be seen from the parameterizations introduced above, the
only surviving contributions  in the forward limit,  are related to  the Compton form factors ${\cal
H}_{1,5}$ and ${\cal G}_{8,9}$ terms. Keeping only these
contributions, one can write
\begin{eqnarray*}
&&\frac{1}{q^2\Delta^2}\sum_i[L^{(0)}{\cal
A}_{(0,i)}]\cdot[\Lambda^{(+,i)}{\cal T}_{(+)}]^* \sim \frac{1}{\xi
(\rho -4) \rho}\times \nonumber\\&& \Big\{ G_1 (16 \xi  (\xi ^2
(\rho -4)-\rho ) (\rho -4) (\xi (\rho -6)+\rho -2) {\cal G}_8^V+16
\xi (\xi (\rho -6)-\rho +2) (\xi
   ^2 (\rho -4)-\rho ) (\rho -4) {\cal G}_9^V
\\
&&-8 (\xi ^2 (\rho -4)-\rho ) (\rho -4) ((\rho -4) \rho +12) {\cal
H}_1^V-16 ((\rho -6) \xi ^2-\rho -2)
   (\xi ^2 (\rho -4)-\rho ) (\rho -4) {\cal H}_5^V)
\\
&&+G_2 (-16 \xi  (\rho -4) ((\rho -4)^2 \xi ^3+(\rho -4) (\rho +2)
\xi ^2-(\rho -8) \rho
   \xi -(\rho -2) \rho ) {\cal G}_8^V-16 \xi  (\rho -4) ((\rho -4)^2 \xi ^3
\\
&&-(\rho -4) (\rho +2) \xi ^2-(\rho -8) \rho  \xi +(\rho -2) \rho )
{\cal G}_9^V+8
   (\xi ^2 (\rho -4)-\rho ) \rho  ((\rho -6) \rho +8) {\cal H}_1^V
\\
&&+16 (\rho -4) (\rho -\xi ^2 (\rho -4))^2 {\cal H}_5^V)+G_3 (16 \xi
   (\xi ^2 (\rho -4)-\rho ) (\rho -4) (\xi  (\rho -4)+\rho ) {\cal G}_8^V
\\
&&+16 \xi  (\xi  (\rho -4)-\rho ) (\xi ^2 (\rho -4)-\rho ) (\rho -4)
{\cal G}_9^V-8(\xi ^2 (\rho -4)-\rho ) \rho  ((\rho -6) \rho +8)
{\cal H}_1^V-16 ((\rho -4)^3 \xi ^4
\\
&&-2 \rho  ((\rho -6) \rho +8) \xi ^2+\rho ^3)
   {\cal H}_5^V) \Big\}+ ....\, .
\end{eqnarray*}
If we now calculate the imaginary part of the above-mentioned terms,
we will obtain the numerator for experimentally accessible single
spin asymmetry parameter \cite{arXiv:0911.0095}.


\section{Conclusions}

In conclusion, we have derived the gauge invariant amplitude for the
deeply virtual Compton scattering off a spin-$1$ hadron. As an
important phenomenological application of this approach, we have
considered the deuteron target and have presented the gauge
invariant DVCS amplitude for the deuteron case. We have also
discussed the simplest kind of asymmetries -- the single spin
asymmetry where the initial lepton has a longitudinal polarization
while all other particles, the initial hadron, the final lepton and
the final hadron, are unpolarized.

\section{Acknowledgements}

The authors would like to thank A.V.~Efremov, G.~Ingelman,
D.~Ivanov, L.~Szymanowski, S.~Wallon for useful discussions and
comments. This work is supported in part by
the DFG project BR2021/6-1,
the RFBR (grants
09-02-01149, 11-02-01454, 12-02-00613) and by the Carl Trygger Foundation (Sweden).

\end{document}